\documentclass[12pt,preprint]{aastex}
\usepackage{amsmath}
\usepackage{amsfonts}
\usepackage{hyperref}

\usepackage{graphicx,epsfig}

\newcounter{theappend}

\newcommand{\be}{\begin{equation}}
\newcommand{\ee}{\end{equation}}
\newcommand{\bea}{\begin{eqnarray}}
\newcommand{\eea}{\end{eqnarray}}
\newcommand{\beaa}{\begin{eqnarray*}}
\newcommand{\eeaa}{\end{eqnarray*}}
\newcommand{\ba}{\begin{array}}
\newcommand{\ea}{\end{array}}
\newcommand{\bi}{\begin{itemize}}
\newcommand{\ei}{\end{itemize}}
\newcommand{\ben}{\begin{enumerate}}
\newcommand{\een}{\end{enumerate}}

\newcommand{\bra}{\langle}
\newcommand{\ket}{\rangle}
\newcommand{\ra}{\rightarrow}

\newcommand{\lb}{\label}
\newcommand{\g}{\gamma}
\newcommand{\G}{\Gamma}
\newcommand{\e}{\epsilon}

\newcommand{\Om}{\Omega}

\newcommand{\sm}{\sigma}

\begin{document}

\title{ {\it Fermi} Gamma-ray Haze \\
via Dark Matter and Millisecond Pulsars 
\bigskip
}

\author{Dmitry Malyshev\altaffilmark{1}, Ilias Cholis,
and Joseph D. Gelfand\altaffilmark{2}}
\email{dm137@nyu.edu, ijc219@nyu.edu, jg168@astro.physics.nyu.edu}

%\author{Ilias Cholis}
% \email{ijc219@nyu.edu}

%\author{Joseph Gelfand}
 %\email{jg168@astro.physics.nyu.edu}

 \affil{CCPP, 4 Washington Place, Meyer Hall of Physics, NYU, New York, NY 10003, USA}

\altaffiltext{1}{On leave of absence from ITEP, Moscow, Russia, B. Cheremushkinskaya 25}
\altaffiltext{2}{NSF Astronomy and Astrophysics Postdoctoral Fellow}
%\altaffiltext{3}{ijc219@nyu.edu}
%\altaffiltext{4}{jg168@astro.physics.nyu.edu}

%\date{\today}

\begin{abstract}

We study possible astrophysical and dark matter (DM) explanations
for the {\it Fermi} gamma-ray haze in the Milky Way halo.
As representatives of various DM models,
we consider DM particles annihilating into $W^+W^-$, $b\bar{b}$, and $e^+e^-$.
In the first two cases,
the prompt gamma-ray emission from DM annihilations
is significant or even dominant
at $E > 10$ GeV,
while inverse Compton scattering (ICS) from annihilating DM products
is insignificant.
For the $e^+e^-$ annihilation mode, we require a boost factor of order 100
to get significant contribution to the gamma-ray haze from ICS photons.
Possible astrophysical sources of high energy particles at high latitudes 
include type Ia supernovae (SNe) and millisecond pulsars (MSPs).
Based on our current understanding of Ia SNe rates,
they do not contribute significantly to gamma-ray flux in the halo of the 
Milky Way.
As the MSP population in the stellar halo of the Milky Way is not well constrained, 
MSPs may be a viable source of gamma-rays at high latitudes
provided that there are $\sim (2 - 6)\times 10^4$ of MSPs 
in the Milky Way stellar halo.
In this case, pulsed gamma-ray emission from MSPs can contribute
to gamma-rays around few GeV's
while the ICS photons from MSP electrons and positrons may be significant 
at all energies in the gamma-ray haze.
The plausibility of such a population of MSPs is discussed.
Consistency with the {\it Wilkinson Microwave Anisotropy Probe}
({\it{WMAP}}) microwave 
haze requires that either a significant fraction of MSP spin-down energy
is converted into $e^+e^-$ flux or
the DM annihilates predominantly into leptons with a boost factor of order 100.
\newpage

\end{abstract}

%\pacs{
%95.35.+d, %Dark matter (stellar, interstellar, galactic, and cosmological)
%96.50.S-, %Cosmic rays
%98.70.Sa %Cosmic rays (including sources, origin, acceleration, and interactions)
%97.60.Gb, %Pulsars 
%}

%\maketitle
%\end{titlepage}

%\newpage

%\tableofcontents

% begin body

Recently, \cite{2010ApJ...717..825D} have found evidence for a $\g$-ray haze
in the halo around the Milky Way Galactic center (GC).
This signal can be a signature of dark matter (DM) annihilation
\citep[e.g.,][]{Zeldovich:1980st, 2008Natur.456...73S, 2009Sci...325..970K}.
The primary purpose of this paper is to look for possible astrophysical sources
of the haze and compare them with DM.
Annihilating DM particles that produce many
prompt $\gamma$'s (via channels such as 
$\chi \chi \rightarrow W^{+}W^{-}$, $ZZ$, $b\bar{b}$, $\tau^{+}\tau^{-}$), 
may significantly contribute or even dominate at photon energies above 10 GeV.
The DM particles that predominantly annihilate into leptons
contribute significantly to the $\g$-ray haze
only if their annihilation cross section is enhanced by a boost factor
of order 100.
This boost factor can be attributed to Sommerfeld enhancement 
\citep{2005PhRvD..71f3528H, 2009PhRvD..79a5014A}
in, e.g., XDM models with annihilation channel 
$\chi\chi \ra \varphi\varphi \ra 2e^+2e^-$
\citep{2007PhRvD..76h3519F, 2009PhRvD..79l3505C, 2009PhRvD..79a5014A}.

A useful discrimination of various sources 
is their total power in the Milky Way halo.
In order to estimate the power of the $\g$-ray haze, let us first calculate it in the 
$\g$-ray haze ``window" 
used by \cite{2010ApJ...717..825D}
to find the spectrum of $\g$-rays in the haze.
This region is
$-15^\circ < l < 15^\circ$ and
$-30^\circ < b < -10^\circ$.
The corresponding solid angle is
$\Om_{\rm haze} = (l_2 - l_1)(\sin b_2 - \sin b_1) \approx 0.17$.
Integrating the $\g$-ray spectrum in Figure 11 of \cite{2010ApJ...717..825D}, we find
\bea
\nonumber
W_{\g - \rm haze} &=&  \Om_{\rm haze} \:4 \pi R_{\odot}^2 \int E \frac{dN_\g}{dE} dE \\ 
\label{g-window}
&\sim& 10^{37}\: {\rm erg\: s^{-1}},
\eea
where $R_{\odot} = 8.5$ kpc is the distance from the GC to the Sun.
The haze is observed within approximately $\theta = 45^\circ$ from the GC. 
The corresponding solid angle $\Om_{\rm tot} = 2\pi(1 - \cos\theta) \approx 1.8$.
Therefore, the total luminosity of this $\g$-ray emission is 
\be
W_{\g\: \rm tot} = \frac{\Om_{\rm tot}}{\Om_{\rm haze}} W_{\g - \rm haze} 
\sim 10^{38}\: {\rm erg\: s^{-1}}
\ee

At first we will discuss possible astrophysical sources of the $\g$-ray haze.
The current star formation rate in the halo of the Milky Way is very small.
Thus the sources of the high energy particles should have a long lifetime.
The two possibilities are type Ia supernovae (SNe) and millisecond pulsars (MSPs).

Let us estimate the output in the high energy electrons from the type Ia SNe.
On average, SNe must produce $\sim 10^{48}$ erg in relativistic electrons
to account for the observed flux of cosmic-ray electrons
\citep{Kobayashi:2003kp}.
The calculations for observed SNe predict similar or smaller power in electrons
\citep{2008A&A...492..695B, 2010ApJ...708..965Z}.
The birth rate of Ia SNe per unit stellar mass in the Galactic halo
can be estimated as 
$(5.3 \pm 1.1) \times 10^{-14}\; {\rm yr}^{-1}\; M_\odot^{-1}$
\citep{2006ApJ...648..868S}.
In order to estimate the mass of the Milky Way stellar halo,
we use the distribution of matter in the disk
and in the halo given by \cite{2008ApJ...673..864J}
and, for the overall normalization,
we use the local stellar density of the thin disk,
$35\: M_\odot\: {\rm pc}^{-2}$ \citep{1989MNRAS.239..571K}.
Therefore, the inferred mass of the Milky Way stellar halo within 20 kpc 
of the GC is 
\be
\lb{HaloMass}
M_{\rm halo} \sim 10^9\: M_\odot
\ee
with an uncertainty of at most a factor of 2.
This gives a Ia SNe rate in the halo of about $5\times 10^{-5}\;{\rm yr}^{-1}$
or $2 \times 10^{-12}\; {\rm s}^{-1}$.
Consequently,
the electron output of the halo Ia SNe is
\be
W_{\rm e^\pm\; Ia} \lesssim 10^{37}\; {\rm {\rm erg\: s^{-1}}}.
\ee
This is insufficient to account for the $\g$-ray haze already based on the 
total output energy
(we need at least $10^{38} \:{\rm erg\: s^{-1}}$).
Provided that the above estimations are good within an order of magnitude,
we conclude that Ia SNe will not contribute a significant number
of electrons and $\g$-rays in the haze region
and we will not consider them in the following.

MSPs are known to emit pulsed $\g$-rays
with a cutoff at a few GeV 
 \citep{2009Sci...325..848A, 2009Sci...325..845A}.
Observations of X-ray nebula around some MSPs show a production of
high energy electrons and positrons
\citep{2003Sci...299.1372S, 2006ApJ...646.1139K},
but the uncertainties in the particle spectrum are rather large.
We will consider two possibilities for the total energy output in $e^+e^-$.
In the first case, we assume that the electron emission from MSPs
is similar to the electron emission from young radio pulsars,
such as the Crab pulsar \citep{1996MNRAS.278..525A}.
In particular, we assume that
a large fraction of the spin-down energy of an MSP
goes into electrons and positrons
with a cut-off in the $e^+e^-$ injection spectrum, $E_{\rm cut} \gtrsim$ 100GeV.
In this case, we demonstrate that
the MSPs may be sufficient to explain $\g$-ray haze at all energies
(although prompt $\g$-ray emission from DM annihilation
may contribute significantly at 
$E \gtrsim 100$ GeV).
In the second case, we assume that $e^+e^-$ emission from MSPs is small,
then the spectrum above 10 GeV requires an additional source,
such as annihilating DM.
At the moment both possibilities for $e^+e^-$ emission
are consistent with observations of  
MSP nebulae \citep{2003Sci...299.1372S}.
The average properties of pulsed $\g$-ray emission
from MSPs for a sample of eight pulsars detected by {\it Fermi} \citep{2009Sci...325..848A}
are presented in Table \ref{msPsrs}.

{\small
\begin{table}[tpb]
\begin{center}
\begin{tabular} {|c|c|c|}
\hline
Index $\G$  &  Cutoff $E_{\rm cut}$ (GeV)  &   ${\rm log}_{\rm 10} {L}$ (${\rm erg\: s^{-1}}$) \\
\hline
$1.5\pm0.4$ & $2.8 \pm 1.9$ &  $33.9 \pm 0.6$\\
\hline
\end{tabular}
\end{center}
\caption{\small 
Some average properties of eight MSPs observed by {\it Fermi}
\citep{2009Sci...325..848A}.
Here we assume that the $\g$-ray spectrum is a power-law with an
exponential cutoff, $F \sim E^{-\G} e^{-E/E_{\rm cut}}$.
The total power in $\g$-rays equals the total spin-down luminosity
times the conversion efficiency ${L}_\g = \eta_\g {L}$.
Assuming the log-normal distribution of ${L}$,
the mean luminosity is  
$\bra  {L} \ket \sim 2\times 10^{34}\; {\rm erg\: s^{-1}}$.
Assuming $\eta_\g \sim 10\%$ we get the 
mean luminosity in pulsed $\g$-rays to be 
$\bra  {L}_\g \ket \sim 2\times 10^{33}\; {\rm erg\: s^{-1}}$.
}
\label{msPsrs}
\end{table}
}

Let us estimate the number of active MSPs in the Milky Way halo.
One of the standard models for the formation of the Galaxy assumes that
the halo is formed before the disk \citep{Tremaine2008}.
The star formation in the halo happens predominantly during the early stages of the evolution.
Based on the location of observed MSPs on the $P \dot{P}$ diagram
\citep{Lorimer2005, 1993ApJ...402..264C},
the length of time they can produce $e^+e^-$ pairs in their magnetosphere is 
on the order of the Hubble time.
Therefore, we expect that any MSPs produced during this period are still active today.
This conclusion is also consistent with the observed characteristic age
of the MSPs \citep{2005AJ....129.1993M, 2007MNRAS.375.1009F}.

Since the direct detection of MSPs in the Milky Way halo is difficult,
we need an indirect way to estimate their number.
An important observation is that the number density of MSPs
depends sensitively on the stellar density.
For instance, globular clusters have much higher stellar density than
the galaxies.
The 47 Tucanae cluster is expected to have about 50 MSPs 
\citep{2009Sci...325..845A}
and a mass of $10^6 \: M_\odot$,
which gives the MSP number to stellar mass ratio of $5 \times 10^{-5}\: M_\odot^{-1}$.
For the Milky Way disk, the local column density of MSPs is estimated to be
$50\:{\rm kpc}^{-2}$ \citep{1997ApJ...482..971C}.
For a local total mass density of
$50\: M_\odot\: {\rm pc}^{-2}$ \citep{1989MNRAS.239..571K},
this gives the MSP number to mass ratio of $10^{-6}\: M_\odot^{-1}$, 
which is 50 times smaller than the corresponding ratio for the 47 Tuc.

The halo of the Milky Way is formed from both the initial
spherical stellar population and via mergers with nearby dwarf galaxies
and stellar clusters 
\citep{2005ApJ...635..931B, 2006ApJ...638..585F, 2009ApJ...702.1058Z}.
Galaxies at high redshift have significantly higher stellar density
than the low redshift galaxies,
for instance, at $z = 2$ the density may be up to 100 times higher
\citep{2009ApJ...697.1290B}.
For comparison, 
the Milky Way stellar density is of order $10^4$ times smaller
than the stellar density of 47 Tuc within half-mass radius 
\citep{1996AJ....112.1487H}.
We expect the MSP number to mass ratio to be determined by the
stellar density at the time of star formation.
Thus the MSP number to stellar mass ratio for the Milky Way halo
should be higher than that for the Milky Way disk but lower than that
in the globular clusters.
For a Milky Way stellar halo mass of $M_{\rm halo} \sim 10^9\: M_\odot$,
the expected number of MSPs is
\be
\lb{MSP2mass}
N_{\rm halo\; MSPs} \sim 1\times 10^3 - 5\times 10^4
\ee
where the lower bound comes from the MSP number to stellar mass ratio
for the Milky Way disk and the upper bound comes from the MSP number to
stellar mass ratio for the 47 Tuc.
Using the parameters from Table \ref{msPsrs}, we find that the total
power emitted by the halo MSPs can be as high as 
\be
W_{\rm halo\; MSPs} \sim 10^{39}\; {\rm erg\: s^{-1}}
\ee
and the power in pulsed $\g$-rays can be $10^{38}$ ${\rm erg\: s^{-1}}$.
This estimation can vary significantly due to uncertainty in the total number of MSPs
in the stellar halo.
It may also depend on the distribution of pulsar properties, e.g., 
the index, the cutoff and the luminosity in pulsed $\g$-rays.
With only eight observed $\g$-ray MSPs,
this distribution cannot be well measured,
but it should be possible in the future with a better observational statistic.
%Also, due to lack of statistics, 
%there may be variation in the average parameters of MSP $\g$-ray
%spectra presented in table \ref{msPsrs}.
Thus, at the moment, we cannot robustly estimate the contribution of MSPs to 
high latitude $\g$-rays.
The main purpose of this work is to point out the possibility of a significant
population of MSPs in the stellar halo of the Milky Way which are a viable source of the 
$\g$-ray haze.
In this paper, we will take the $\g$-ray luminosity of MSPs in Table \ref{msPsrs} as
a reference value.
Then, for some particular DM models,
we will be able to estimate the number of MSPs in the stellar halo from
the $\g$-ray haze data.

MSPs are also continuously created in the Galactic disk.
Every pulsar acquires a kick velocity at the birth.
For regular pulsars the mean kick velocity is about 400 ${\rm km\: s^{-1}}$
\citep{2006ApJ...643..332F}.
MSPs are usually found in binaries. 
The center of mass velocities of the 
binary systems have typical values $\sim 80$ ${\rm km\: s^{-1}}$
\citep{1997ApJ...482..971C}.
\cite{1998MNRAS.295..743L} obtained
the average velocity $\sim 130$ ${\rm km\: s^{-1}}$ for their sample of MSPs.
These velocities are not sufficient for the majority of MSPs to 
overcome the gravitational potential of the disk
\citep{1997ApJ...482..971C, 2007ApJ...671..713S}.
Although the disk MSPs contribute significantly 
to the diffuse $\g$-ray background 
\citep{2007ApJ...671..713S, 2010JCAP...01..005F},
the corresponding flux of $\g$-rays has a disk-like morphology 
which is different from an egg-like shape of the 
$\g$-ray haze \citep{2010ApJ...717..825D}.
Consequently, we need an additional population
of MSPs that stretches to higher latitudes.

The distribution of the stellar mass in the halo falls approximately
as $r^{-3}$
\citep{2008ApJ...673..864J, 2008ApJ...680..295B}.
In order to get a smooth distribution in the Galactic Center we introduce a cutoff
radius $R_{\rm c}$, and take the following profile of MSPs
\be
\lb{mass_profile}
\rho_{\rm MSPs} \sim \frac{1}{(r+R_{\rm c})^3},
\ee
where we will take $R_{\rm c} = 2$ kpc as our reference value.
Physically, at $R_{\rm c}$ the power-law distribution of the matter in the halo
breaks down and merges to a Gaussian or exponential distribution 
near the GC \citep{1995ApJ...445..716D}.
Since we are interested in high latitude fluxes, the precise shape of the 
matter distribution near the GC is not important
and the approximation in Equation (\ref{mass_profile}) is sufficient.

In our calculations, we will use the following source function for
$e^\pm$ and $\g$ emission from the MSPs
(the parametric form of the source functions
for $e^\pm$ and for $\g$ are the same but the index and the 
cutoff are different)
\be
Q_{\rm MSPs}(r,\: E) = Q_{0} \frac{1}{(r+R_{\rm c})^3} 
\left(\frac{E}{E_{\rm cut}}\right)^{-n} 
{\rm exp}\left(-\frac{E}{E_{\rm cut}}\right),
\ee
where for the $\g$-ray spectrum we will
use the parameters from Table \ref{msPsrs},
$n = 1.5 \pm 0.4$ and $E_{\rm cut} = 2.8 \pm 1.9$.
Motivated by possible similarities between the electron spectrum from
MSPs and younger radio pulsars,
we use the following parameters to describe the spectrum of electrons:
$n = 1.5 \pm 0.5$ and $E_{\rm cut}$ equal to a few hundred GeV
(see, e.g., \cite{1996MNRAS.278..525A} for the $e^+e^-$ spectrum 
in the Crab Nebula).
The total energy output is proportional to the total stellar mass of the halo
in Equation (\ref{HaloMass}) times the MSP number to stellar mass 
ratio in Equation (\ref{MSP2mass}).

Annihilating DM provides a source for prompt $\g$-rays and $e^+e^-$
of the form
\be
\lb{DM_flux_annih}
Q_{\rm DM} (r, E) = 
\frac{1}{2}\frac{\rho_{\rm DM}^2}{M_{\rm DM}^2} \bra \sigma v \ket_0 \frac{dN}{dE},
\ee
where $\rho_{\rm DM}^2 / M_{\rm DM}^2$ is the DM number density,
$\langle\sigma v\rangle_0 = 3.0 \times 10^{-26}\text{cm}^{3}\text{s}^{-1}$
is the thermally averaged annihilation cross section at freeze out,
and ${dN}/{dE}$ is the differential energy spectrum 
of either $e^\pm$ or $\g$ produced in a single annihilation event.
In general, the annihilation rate can be enhanced by a boost factor
which we denote as BF.
For the calculations we use the Einasto profile of DM
\citep{2005ApJ...624L..85M}:
\be
\lb{EinProf}
\rho_{\rm DM}(r) = \rho_{\rm DM_{0}} 
{\rm exp} \left(
-\frac{2}{\alpha}
\frac{r^{\alpha}-R_{\odot}^{\alpha}}{R_{-2}^{\alpha}}
\right),
\ee
where 
$R_{-2}=25$ kpc, $\alpha=0.17$,
and $R_{\odot}=8.5$ kpc.
For the local DM density, we take 
$\rho_{\rm DM_{0}} = 0.4\:\text{GeV} \text{cm}^{-3}$ \citep{2010JCAP...08..004C}.
\cite{2005ApJ...624L..85M} motivated the use of 
this profile by the DM $N$-body simulations.
The rotation curves in the inner part of the Galaxy
are dominated by the luminous matter and do not constrain
significantly the 
DM distribution
(e.g., 
\cite{1998MNRAS.294..429D},
see also 
\cite{2007MNRAS.378...41S}
for a recent fit with a less cuspy profile).
In any case, 
the spectrum of photons in the $\g$-ray haze window
is not sensitive to the choice of DM profile.

The total power released by annihilating DM is
\be
W_{\rm DM} = \frac{1}{2} \int \frac{\rho_{\rm DM}^2}{M_{\rm DM}^2} \bra \sigma v \ket_0 2 M_{\rm DM}\;
4\pi r^2 dr.
\ee
For a DM particle with the mass $M_{\rm DM} = 300$ GeV
and the rest of the parameters described above, 
the total power from annihilating DM
within 20 kpc from GC is
\be
W_{\rm DM} \sim 2 \times 10^{37}\;{\rm erg\: s^{-1}},
\ee
which is about 5 times smaller than the 
total power in the $\g$-ray haze.
Based on this crude estimate we already expect that 
without boost factors this DM model
will not explain the $\g$-ray haze at all energies 
but it can be significant at some energies provided that the 
production of high energy photons is efficient.

For the propagation of $e^{\pm}$ in the interstellar medium we use GALPROP 
\citep{2007ARNPS..57..285S, 1999PhRvD..60f3003M}, 
with the interstellar radiation field model of
\cite{2005ICRC....4...77P}. 
We assume the following magnetic field model
\be 
\label{eq:B-field}
B(R, z)=B_{0}\:{\rm exp}\left(\frac{R_{\odot} - R}{R_{c}}\right)
{\rm exp}\left(-\frac{z}{z_{c}}\right),
\ee
where $B_{0}=5\: \mu$G is the value for the total (combined random and large scale ``ordered'') local magnetic field.
$R_{\rm c}=4.5$ kpc and $z_{c}=2.0$ kpc are the characteristic scales along 
$R$ and $z$ directions, respectively, and
$R_{\odot}=8.5$ kpc.

\begin{figure}[t] %[htbp] here, top, bottom, page
\begin{center}
\epsfig{figure = 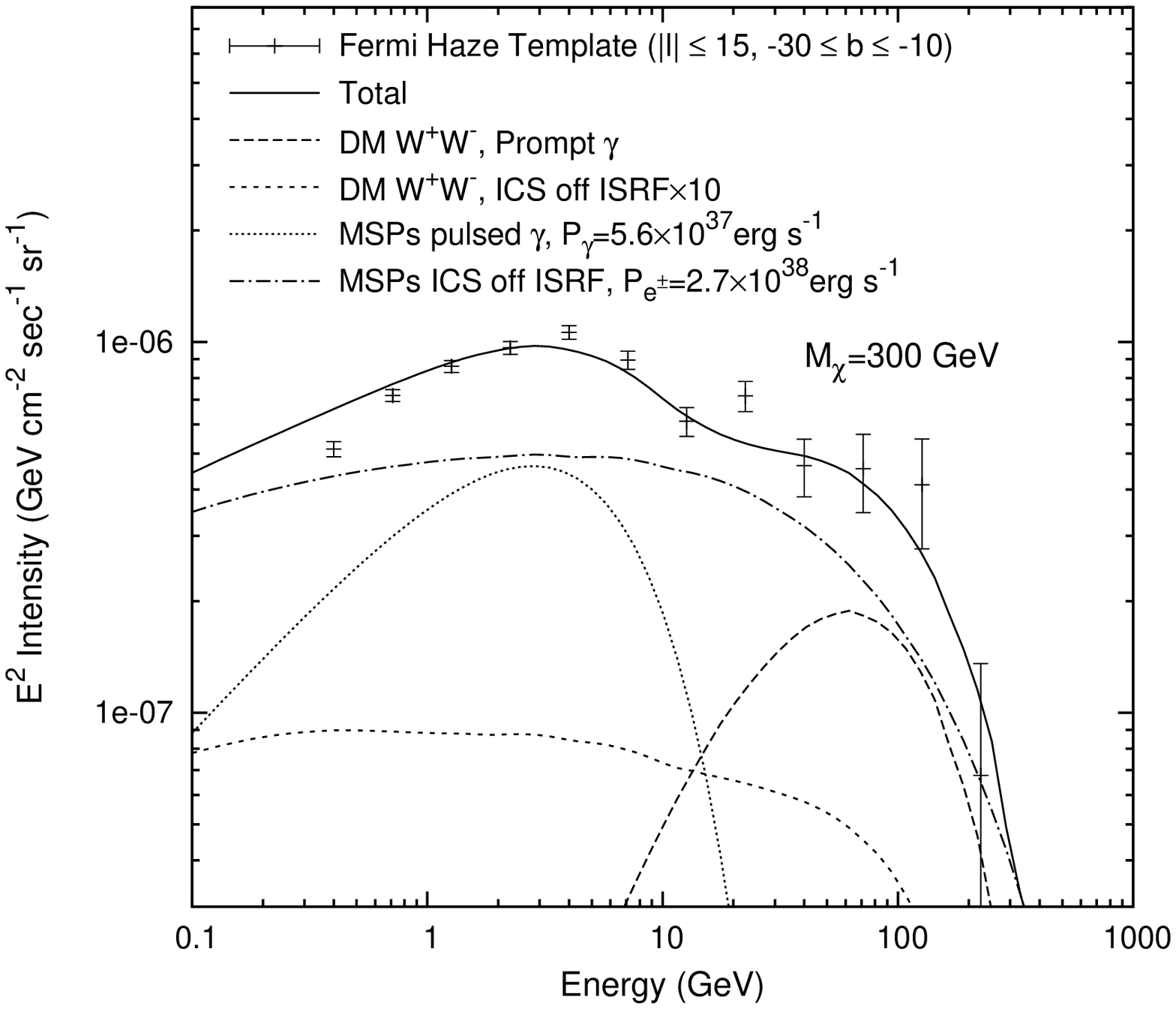,scale=0.45}\;\;\;
\epsfig{figure = 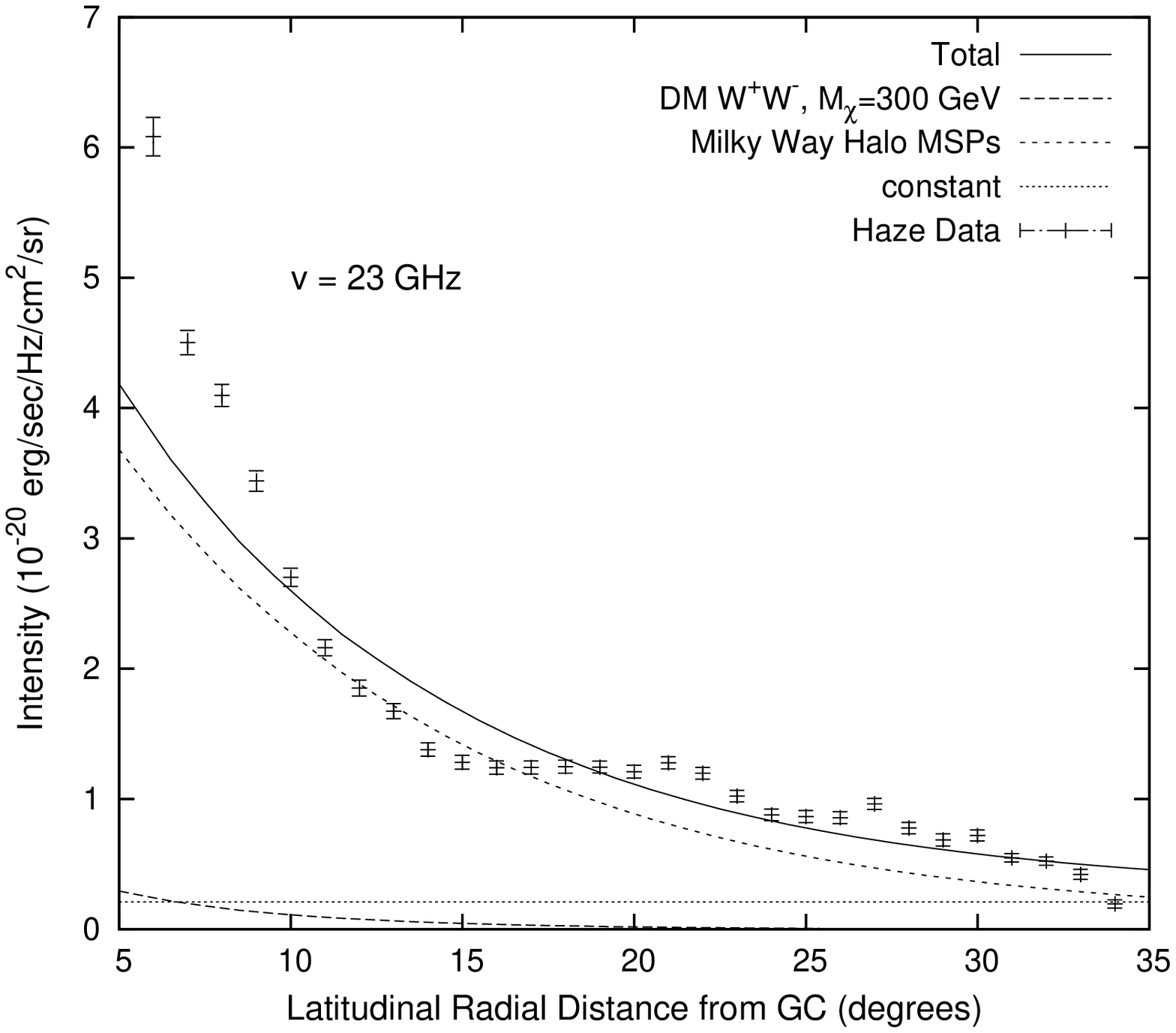,scale=0.45}
\end{center}
\vspace{-8mm}
\noindent
\caption{\small 
Left: contribution of pulsed $\g$-ray emission from MSPs,
prompt $\g$-ray emission from annihilating DM, and 
ICS off MSP and DM electrons to the $\g$-rays
haze spectrum.
Right: corresponding contribution of the synchrotron radiation from the electrons
to the microwave haze at 23 GHz.
The parameters for the pulsed $\g$-ray emission are $n_\g = 1.3$, 
and $E_{\rm cut_{\gamma}} = 4$ GeV.
The MSP $e^+e^-$ injection spectrum parameters are $n_e = 1.3$, and 
$E_{\rm cut_{e}} = 300$ GeV.
The DM has a mass $M_{\rm DM} = 300$ GeV and annihilates into $W^+W^-$
with
$\langle\sigma v\rangle_0 = 3.0 \times 10^{-26}\text{cm}^{3}\text{s}^{-1}$.
The spatial distribution of MSPs and DM is discussed in the text.
We use $R_{\rm c} = 2$ kpc for the distribution of MSPs.
The total power in pulsed $\g$-rays and in $e^+e^-$ emission from MSPs is
$W_\g = 5.6 \times 10^{37}$ ${\rm erg\: s^{-1}}$ and $W_{e^\pm} = 2.7 \times 10^{38}$ ${\rm erg\: s^{-1}}$, respectively.
For a mean ${L}_{\rm MSP} \sim 2\times 10^{34}$ ${\rm erg\: s^{-1}}$, it corresponds to about $3\times 10^4$
halo MSPs with average
conversion efficiencies $\eta_\g \approx 0.1$ and $\eta_{\e^\pm} \approx 0.5$.
}
\label{GammaHaze}
\vspace{1mm}
\end{figure}

Our main result is shown in Figure \ref{GammaHaze}.
In order to explain the $\g$-ray haze,
we add the direct emission of pulsed $\g$-rays from MSPs,
the ICS photons from MSP and DM electrons,
and prompt $\g$-ray emission from annihilating DM.
The prompt $\g$-ray emission from annihilating DM becomes comparable to 
ICS photons from MSP electrons around 100 GeV.
The bump in the $\g$-ray haze spectrum around 4 GeV is naturally explained with the 
pulsed $\g$-ray emission from MSPs.
%The ICS photons have a very flat spectrum and do not explain this feature. 
We also compare the synchrotron radiation with the microwave haze data
\citep{2004ApJ...614..186F}.
We add the synchrotron from MSP and DM electrons together with a constant
a priori unknown offset.
Our model seems to give synchrotron radiation that is less cuspy than 
the microwave haze data.
The deviation happens below $10^\circ$ which corresponds to a distance of about 1.5 kpc from the GC
where our mass distribution in Equation (\ref{mass_profile}) may be invalid.
We also expect some contribution from 
sources in the bulge which may provide cuspier latitudinal 
profile of synchrotron radiation than the one presented in Figure~\ref{GammaHaze}.

In this example, we need about $3\times 10^4$ MSPs in the Milky Way halo
which corresponds to $(1 - 2) \times 10^3$ MSPs in the $\g$-ray haze ``window".
The question is whether it is possible to distinguish between the $\g$-ray flux from MSPs 
and the diffuse ICS photons.
The flux from an average MSP with pulsed $\g$-ray luminosity ${L}_\g \sim 10^{33}\: {\rm erg\: s^{-1}}$
at a distance of 8.5 kpc
from the Earth is $\lesssim 10^{-12} \:{\rm erg}\:{\rm s}^{-1} {\rm cm}^{-2}$,
which is less than the {\it Fermi} sensitivity for a detection of a point source
with $8 \sigma$ significance
even after a year of observation.%
\footnote{
{\it Fermi}-LAT performance:
\url{http://www-glast.slac.stanford.edu/software/IS/glast_lat_performance.htm}
}
One can also use more sophisticated methods to distinguish between
the point sources and
the diffuse background without point source identification
\citep{2010MNRAS.405.1777S}.
The problem is that the expected typical separation between MSPs
in the $\g$-ray haze ``window" is comparable to the {\it Fermi}-LAT point spread function 
at energies $< 10$ GeV 
\citep{2009arXiv0907.0626R}.
Consequently, it seems implausible that one will be able to distinguish between the
$\g$-ray emission from MSPs and the diffuse ICS photons
using available $\g$-ray observations.

\begin{figure}[t] %[htbp] here, top, bottom, page
\begin{center}
\epsfig{figure = 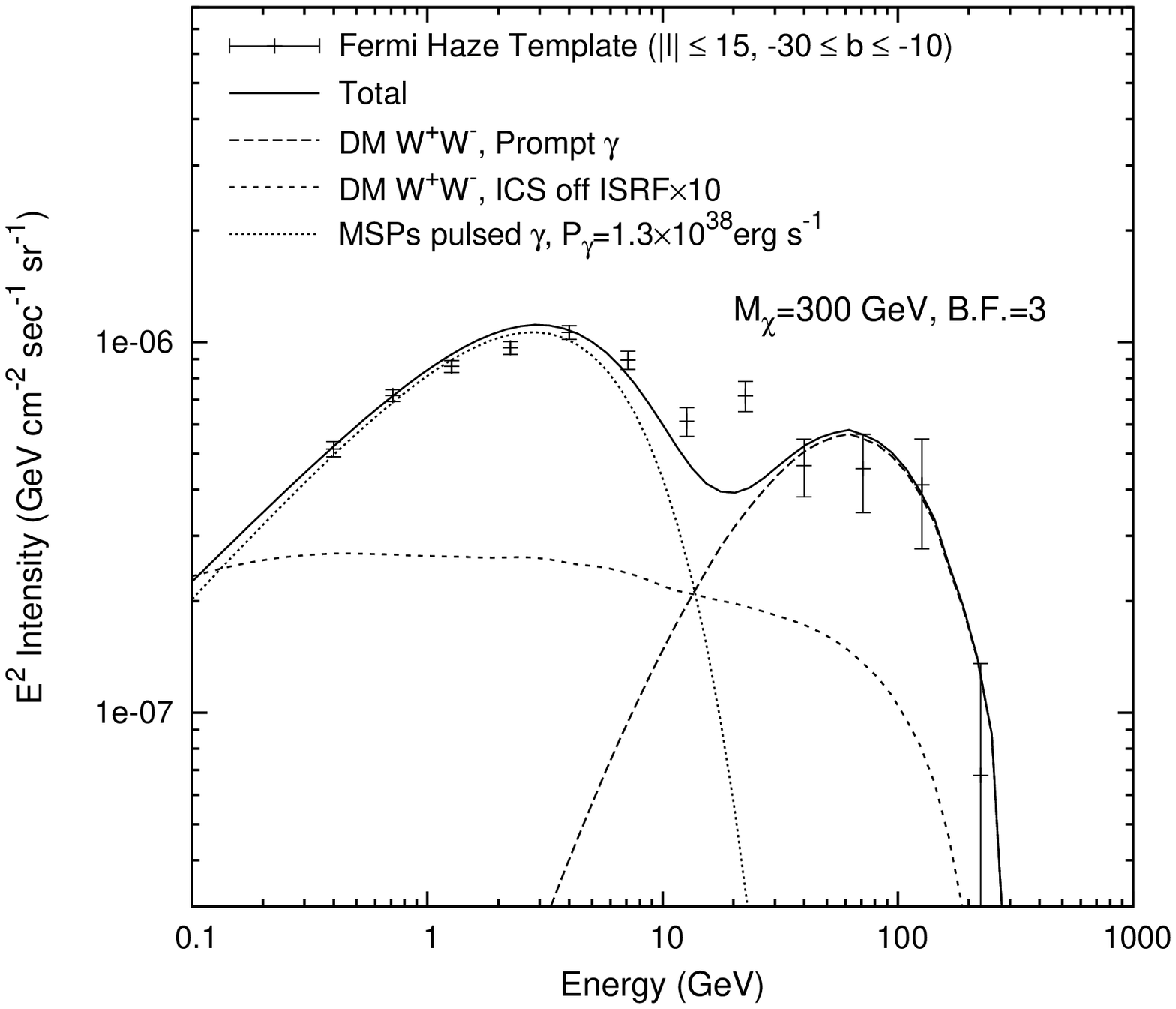,scale=0.45}\;\;\;
\epsfig{figure = 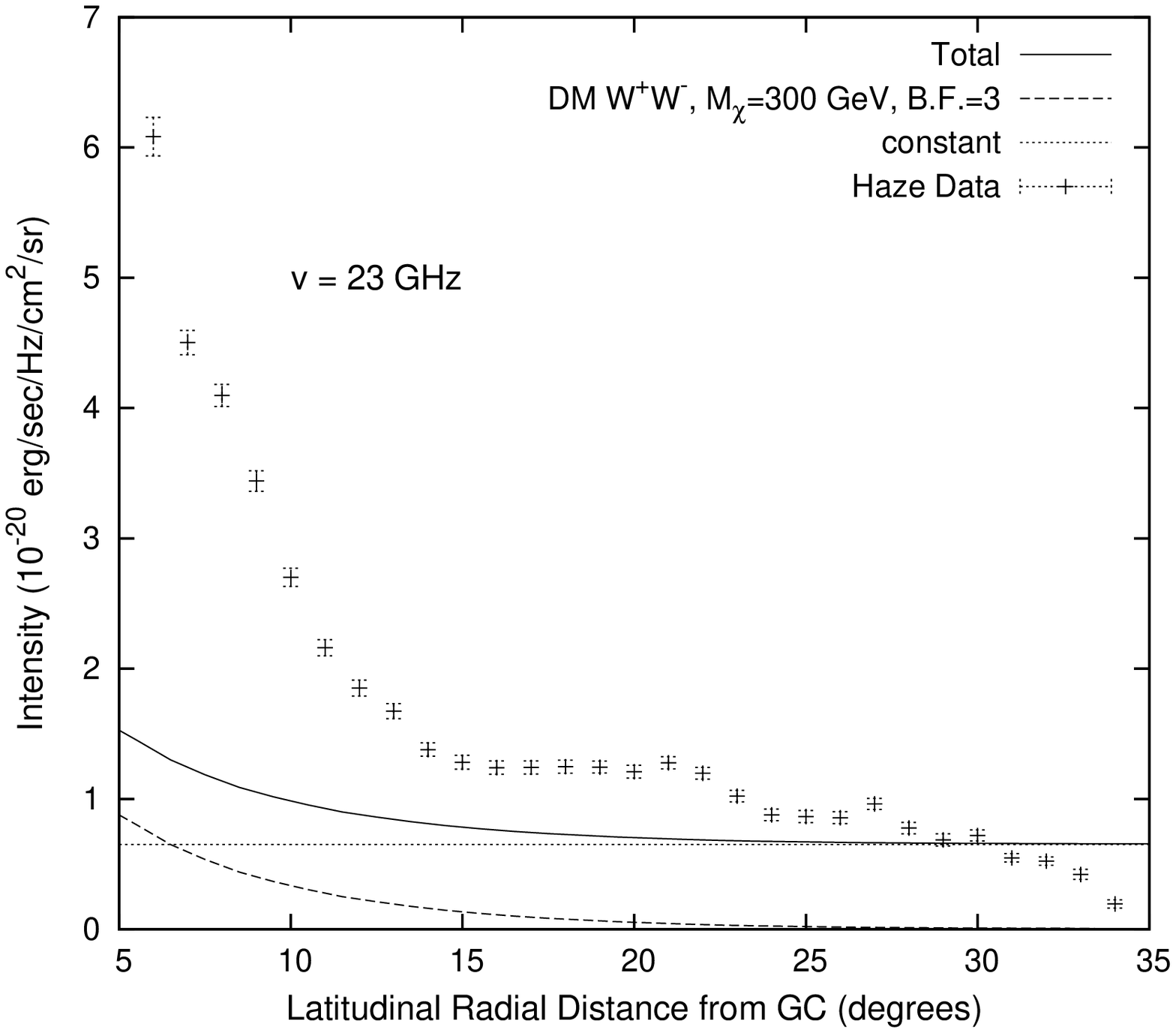,scale=0.45}
\end{center}
\vspace{-8mm}
\noindent
\caption{\small 
Same as in Figure \ref{GammaHaze}, but without MSP electrons.
For the $\g$-ray haze, 
we need about 2.3 more power in pulsed $\g$-ray emission and 
a boost factor of 3 for the DM relative to Figure \ref{GammaHaze}.
For a mean luminosity in pulsed $\g$-rays
${L}_{\g\:\rm MSP} \sim 2\times 10^{33}$ ${\rm erg\: s^{-1}}$, it corresponds to 
a population of about $6\times 10^4$ MSPs in the Milky Way halo
which is slightly higher than the upper bound in Equation (\ref{MSP2mass}).
}
\label{GammaHazeNoMSPe}
\vspace{1mm}
\end{figure}

Let us argue now that a population of $3\times 10^4$ MSPs in the Milky Way halo
is not inconsistent with the properties of currently known MSPs.
Since the luminosity of MSPs is generally very low,
most of the observed MSPs are within $\sim 1 - 2$ kpc from the Earth.
Thus, we cannot directly observe the halo MSPs at large heights above the disk 
and the only parameter that can effectively distinguish a halo MSP
from a disk MSP is the velocity relative to the Sun.
Based on the force in the vertical direction at $R = R_\odot$, $z = 1.1$ kpc
\citep{1991ApJ...367L...9K},
$K_{z,1.1} = 2 G \times (71 \pm 6)M_\odot {\rm pc}^{-2}$,
we can estimate the escape velocity $v_{\rm esc} \sim 100$ ${\rm km\: s^{-1}}$.
Pulsars with the transverse velocity $v_{\rm tr} \gg v_{\rm esc}$ can be attributed 
to the halo population.

According to the ATNF catalog \citep{2005AJ....129.1993M},
there are about three MSPs with $v_{\rm tr} \gtrsim 200$ ${\rm km\: s^{-1}}$ at 
a distance $r \sim 1$ kpc from the Earth
(e.g., B1957+20, J1909-3744, and B1257+12).
This is consistent with the 
local number density that follows from the distribution in Equation
(\ref{mass_profile}) $n_{\rm halo\:MSP} \sim 3\;{\rm kpc}^{-3}$.
One should also take into account that not all local MSPs can be
observed due to beaming of the radio emission.
The local MSP number density that we need is significantly larger 
than the one estimated by \cite{1997ApJ...482..971C},
$n_{\rm h} < 0.4\:{\rm kpc}^{-3}$ at $90\%$ confidence.
We believe that this discrepancy may be due to the small 
velocity dispersion $\sm_v = 53$ ${\rm km\: s^{-1}}$ 
in the sample of MSPs used by \cite{1997ApJ...482..971C}.
This velocity dispersion implies that the probability to have $v_{\rm tr} > 200$
${\rm km\: s^{-1}}$ is less than $10^{-3}$
which is inconsistent with a few MSPs out of about 100
in the ATNF catalog that have
$v_{\rm tr} > 200$.
The analysis of \cite{1998MNRAS.295..743L} gives 
$\sm_v = (80 \pm 20)$ ${\rm km\: s^{-1}}$
which corresponds to the probability 
$P(v_{\rm tr} > 200\:{\rm km\: s^{-1}})\approx 4\%$.
This probability is consistent with both the currently observed MSPs and the
local halo MSP number density $n_{\rm halo\:MSP} \sim 3\;{\rm kpc}^{-3}$
relative to the disk MSP number density 
$n_{\rm disk\:MSP} \sim 40\;{\rm kpc}^{-3}$
\citep{1997ApJ...482..971C}.

In Figure \ref{GammaHazeNoMSPe},
we consider the case that MSPs do not emit enough electrons
and fit the $\g$-ray haze data with the pulsed emission from MSPs
and with DM annihilation products.
In this case, we need a rather mild boost factor BF = 3.
For the microwave haze data we need a significantly larger constant offset
than in Figure \ref{GammaHaze},
and the shape of the microwave haze is not fit nicely due to the small production rate 
of $e^+e^-$ by annihilating DM in this case.

In both of the cases considered above the $e^+e^-$ emission 
from MSPs and DM is too small to significantly contribute to local $e^+e^-$ fluxes.
In particular, we need to assume that the rising positron fraction in {\it PAMELA} experiment
\citep{2009Natur.458..607A}
is due to a different astrophysical source, such as young radio pulsars 
\citep{2009JCAP...01..025H, 2009PhRvL.103e1101Y, 2008arXiv0812.4457P, 2009PhRvD..80f3005M}.

\begin{figure}[t] %[htbp] here, top, bottom, page
\begin{center}
\epsfig{figure = 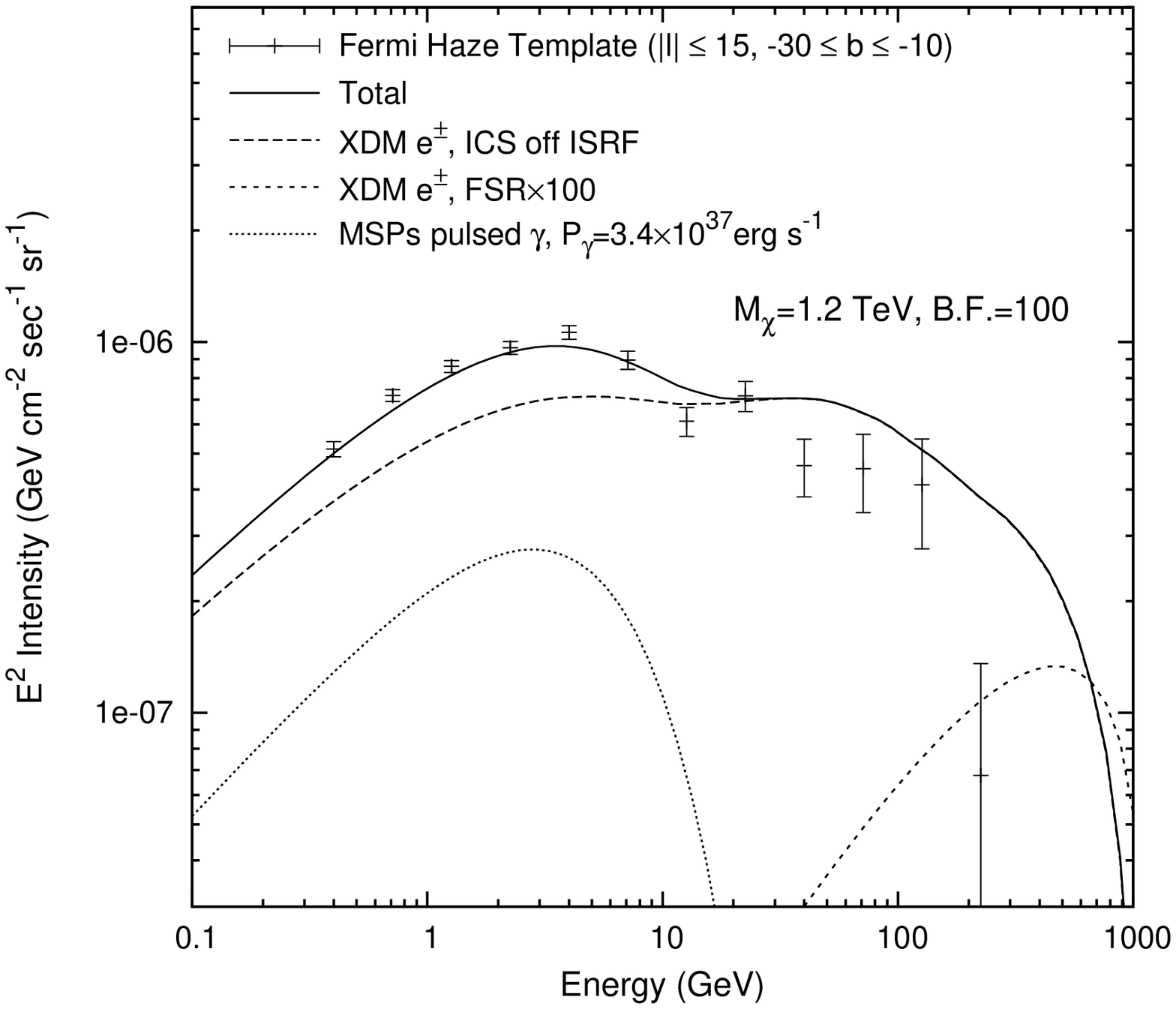,scale=0.45}\;\;\;
\epsfig{figure = 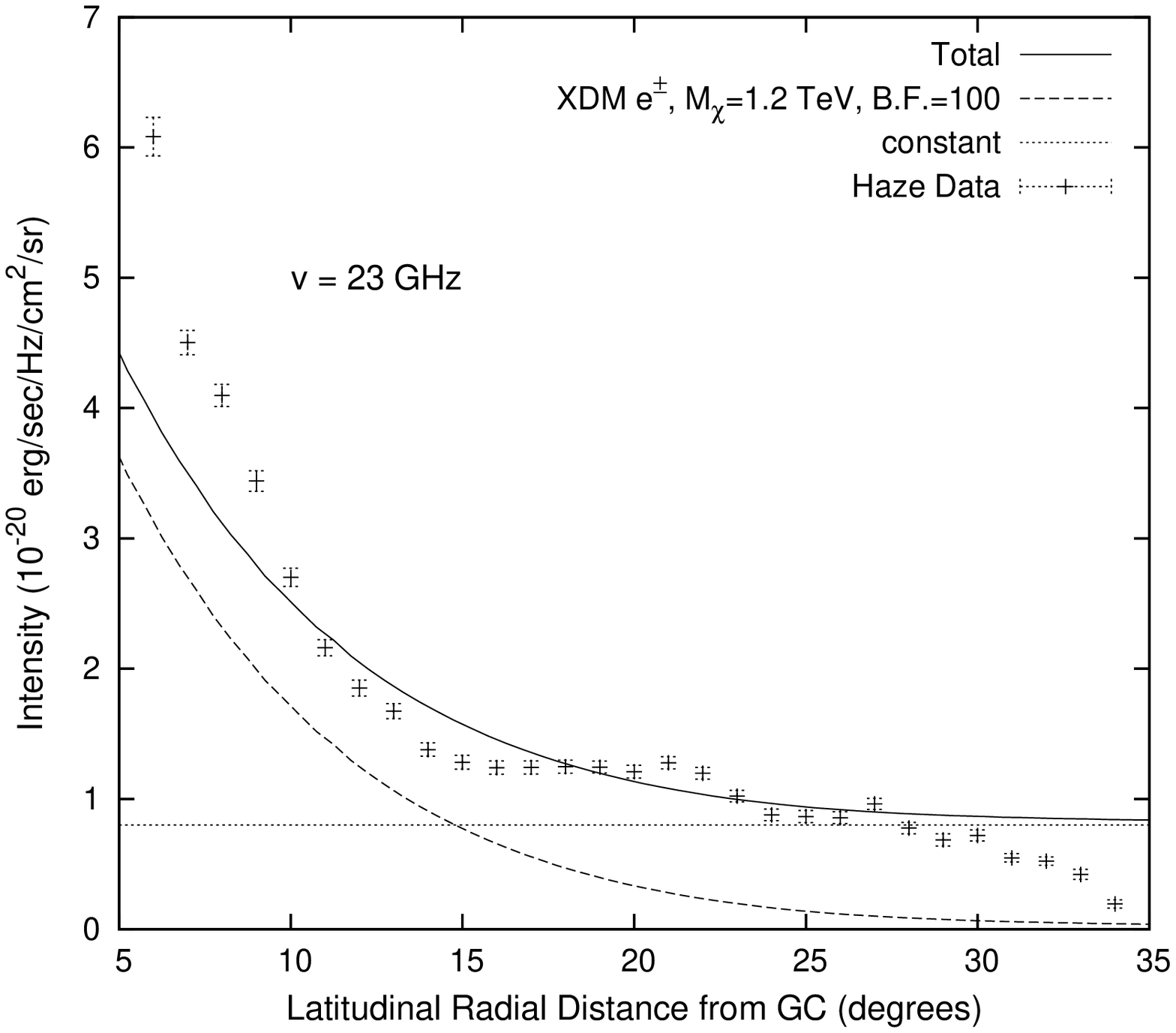,scale=0.45}
\end{center}
\vspace{-8mm}
\noindent
\caption{\small 
Pulsed $\g$-ray emission from MSPs and 
DM with $e^+e^-$ annihilation mode.
This DM model was
used by \cite{2009PhRvD..80l3518C} to fit the local $e^+e^-$ cosmic-ray experiments.
The spectrum of ICS photons is too flat to reproduce the bump around 4 GeV.
As a result, a contribution of $\g$-rays from a population of about $2\times 10^4$ 
MSPs in the Milky Way halo is necessary.
}
\label{XDMandMSP}
\vspace{1mm}
\end{figure}

In the case when $e^+e^-$ emission from MSPs is small, one can also 
consider DM particles that predominantly annihilate into leptons with large boost factors.
As an example,
we study a DM model \citep{2009PhRvD..79a5014A}
used to explain the local $e^+e^-$ fluxes
\citep{2009PhRvD..80l3518C}.
%measured by \citep{2009Natur.458..607A, 2008Natur.456..362C, 
%2009PhRvL.102r1101A, 2008PhRvL.101z1104A}.
The corresponding $\g$-rays signal is shown in Figure \ref{XDMandMSP}.
The ICS photons have a flat spectrum and cannot explain the bump
around 4 GeV which can be produced by pulsed $\g$-ray emission from
about $20\,000$ MSPs in the Milky Way halo.
Another possibility is to consider a model with more than one species of DM particles 
\citep[e.g.,][]{2009arXiv0911.4954C}.

\begin{figure}[t] %[htbp] here, top, bottom, page
\begin{center}
\epsfig{figure = 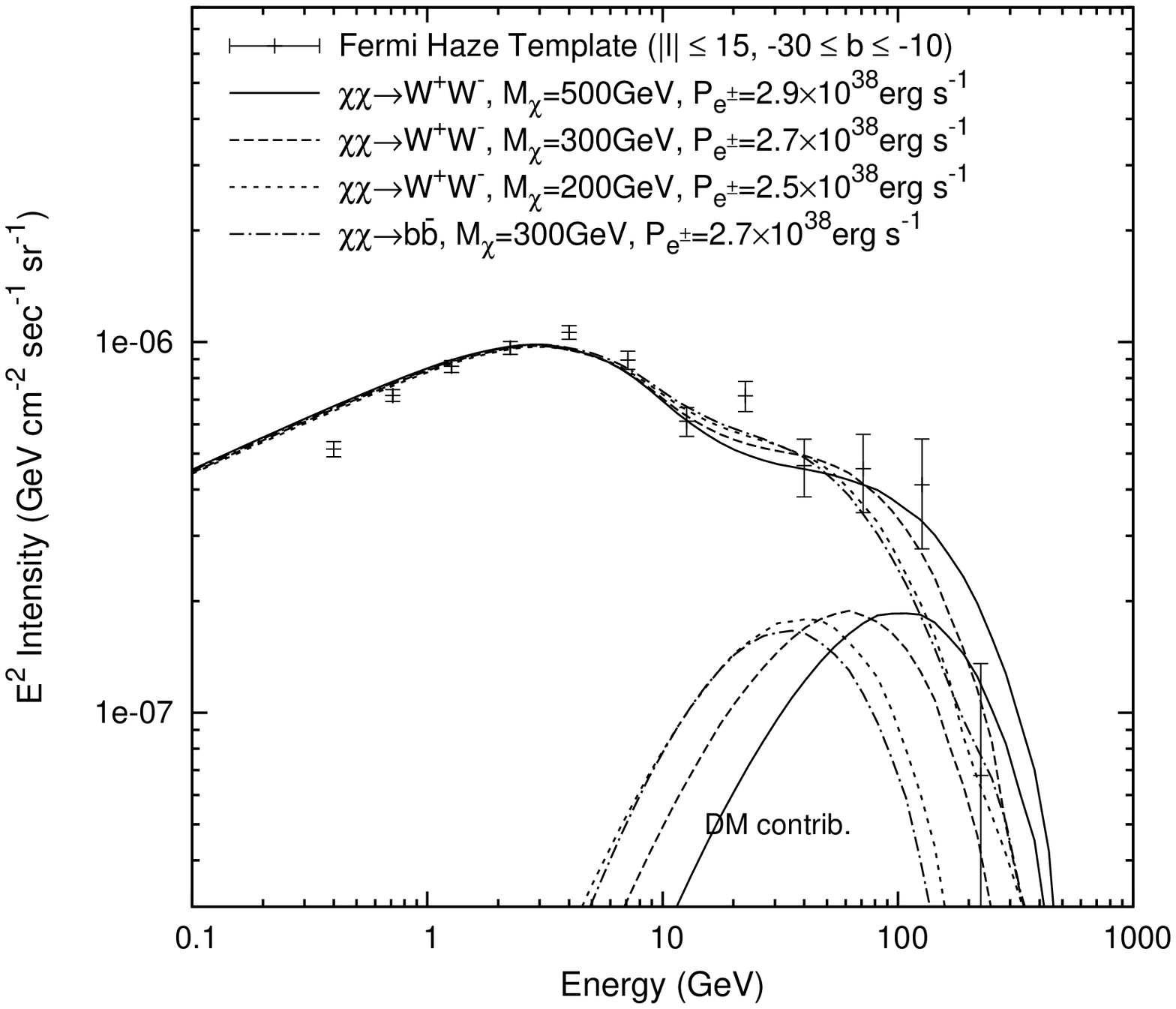, scale=0.45}
\epsfig{figure = 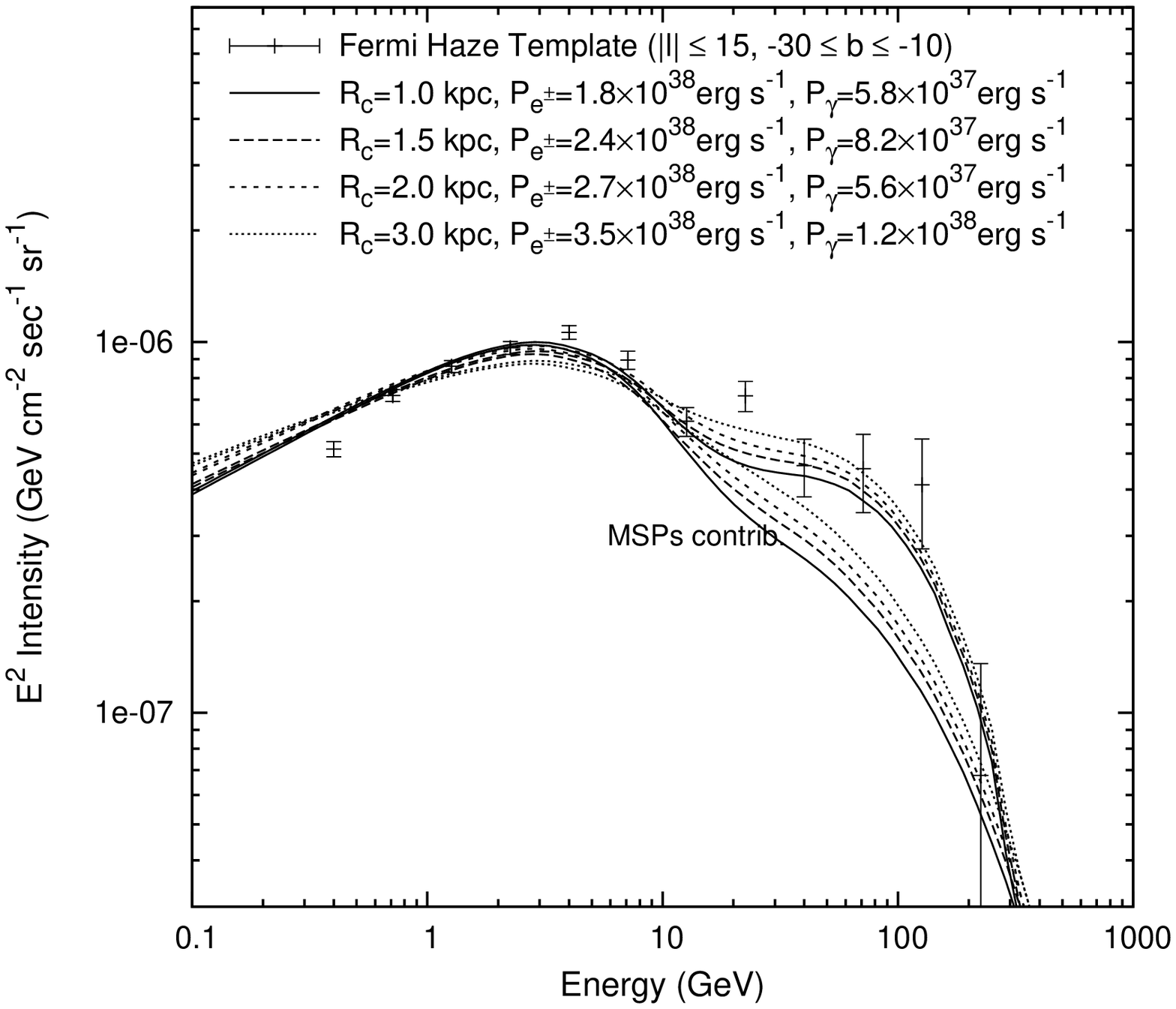, scale=0.45}\\
\epsfig{figure = 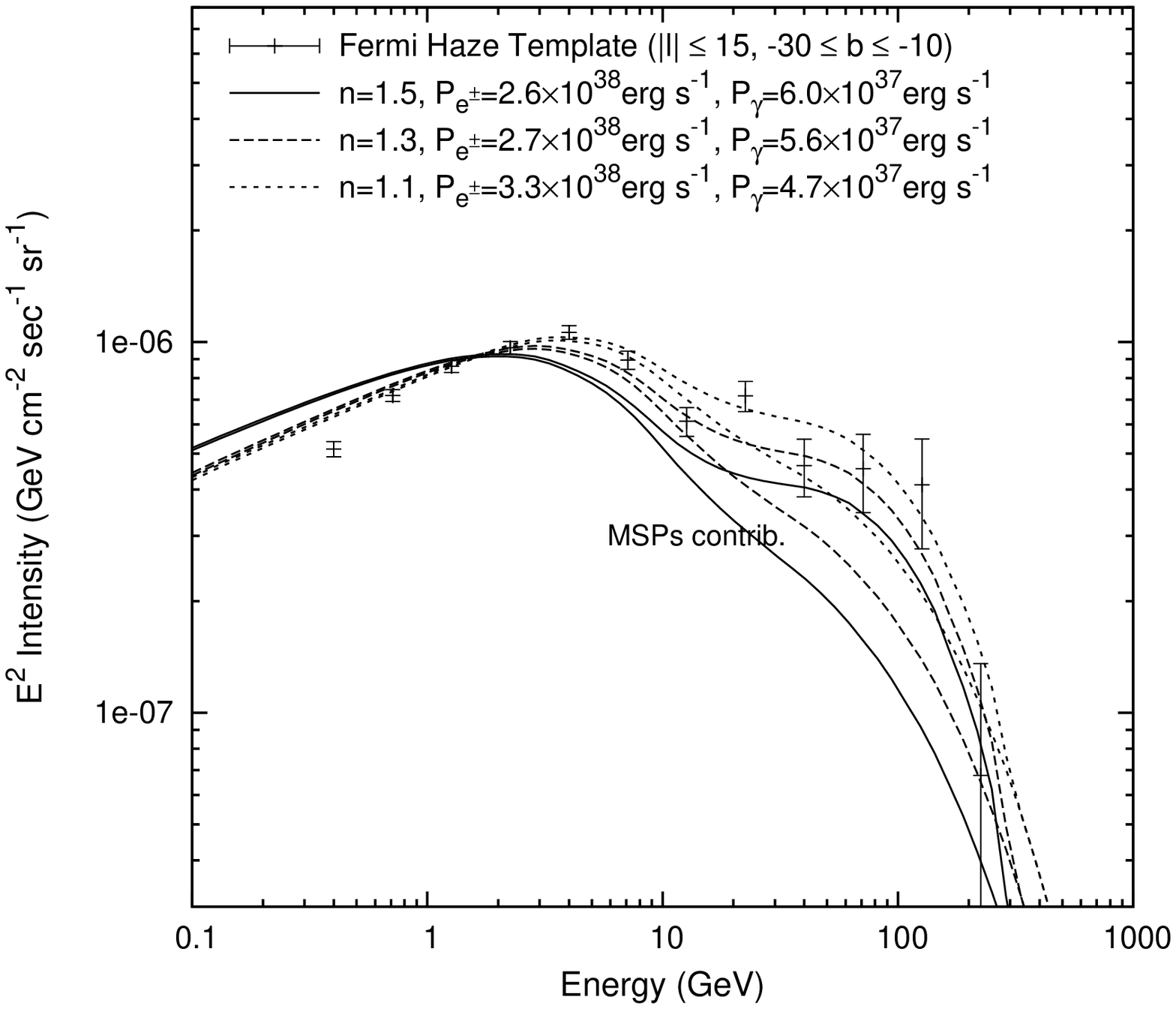, scale=0.45}
\epsfig{figure = 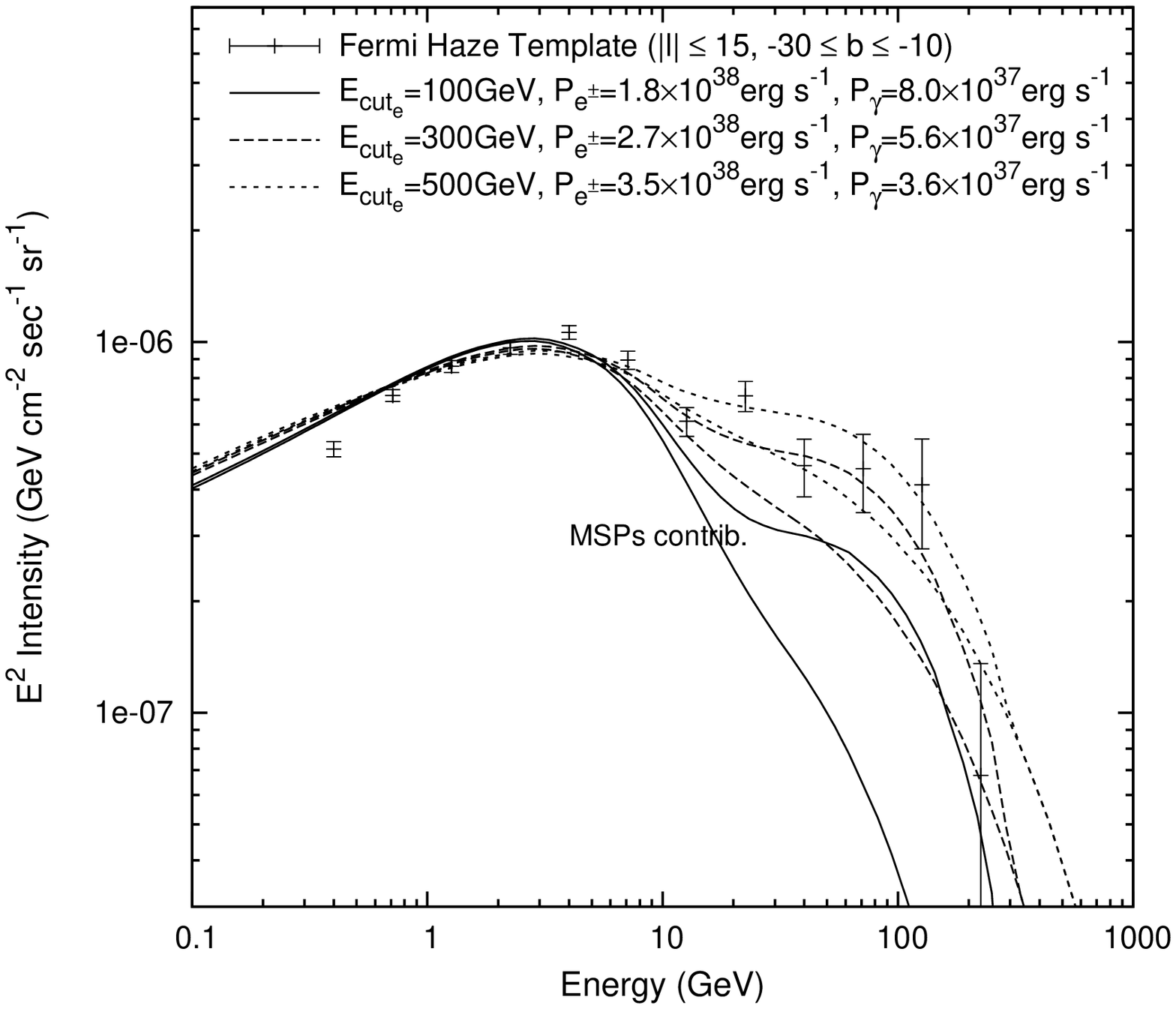, scale=0.45}
\end{center}
\vspace{-8mm}
\noindent
\caption{\small 
Fits to $\g$-ray haze data from Milky Way halo MSPs and DM with varying parameters. 
In each plot we present both the combined contribution from MSPs and DM and 
the contribution of the component whose properties we vary in each case. 
The reference model is as shown in Figure~\ref{GammaHaze}. 
DM has a mass of 300 GeV and annihilates via $\chi\chi \rightarrow W^{+}W^{-}$ 
with $\bra \sigma v \ket = 3\times10^{-26}$ ${\rm cm}^{3}\: {\rm s}^{-1}$.
The MSPs have distribution profile $\rho_{MSP} \sim 1/(r+R_{\rm c})^{3}$, where 
$R_{\rm c}=2$kpc. 
The MSP $e^{+}e^{-}$ reference injection parameters are $n_{e}=1.3$ and
$E_{\rm cut_{e}}=300$ GeV.
The pulsed $\gamma$-ray emission parameters are $n_{\gamma}=1.3$
and $E_{\rm cut_{\gamma}}=4$ GeV. 
Upper left: varying DM mass $M_{\rm DM} = 200,\: 300,\: 500$ GeV for $\chi\chi \rightarrow W^{+}W^{-}$ 
and $M_{\chi}=300$ GeV for  $\chi\chi \rightarrow b \bar{b}$. 
Upper right: varying scaling radius $R_{\rm c}$ of MSPs distribution profile. 
Lower left: varying MSPs electron injection (and $\gamma$-ray) index ($n=n_{e}=n_{\gamma}$). Lower right: varying MSPs electron injection cutoff $E_{e_{0}}$.
On each plot, there are two lines of the same type: 
two solid lines, two dashed lines, two dotted lines, etc.
The lower one has only MSPs or only DM
contribution, while the upper one has MSPs plus DM contributions.
}
\label{VaryParsFermi}
\vspace{1mm}
\end{figure}

\begin{figure}[t] %[htbp] here, top, bottom, page
\begin{center}
\epsfig{figure = 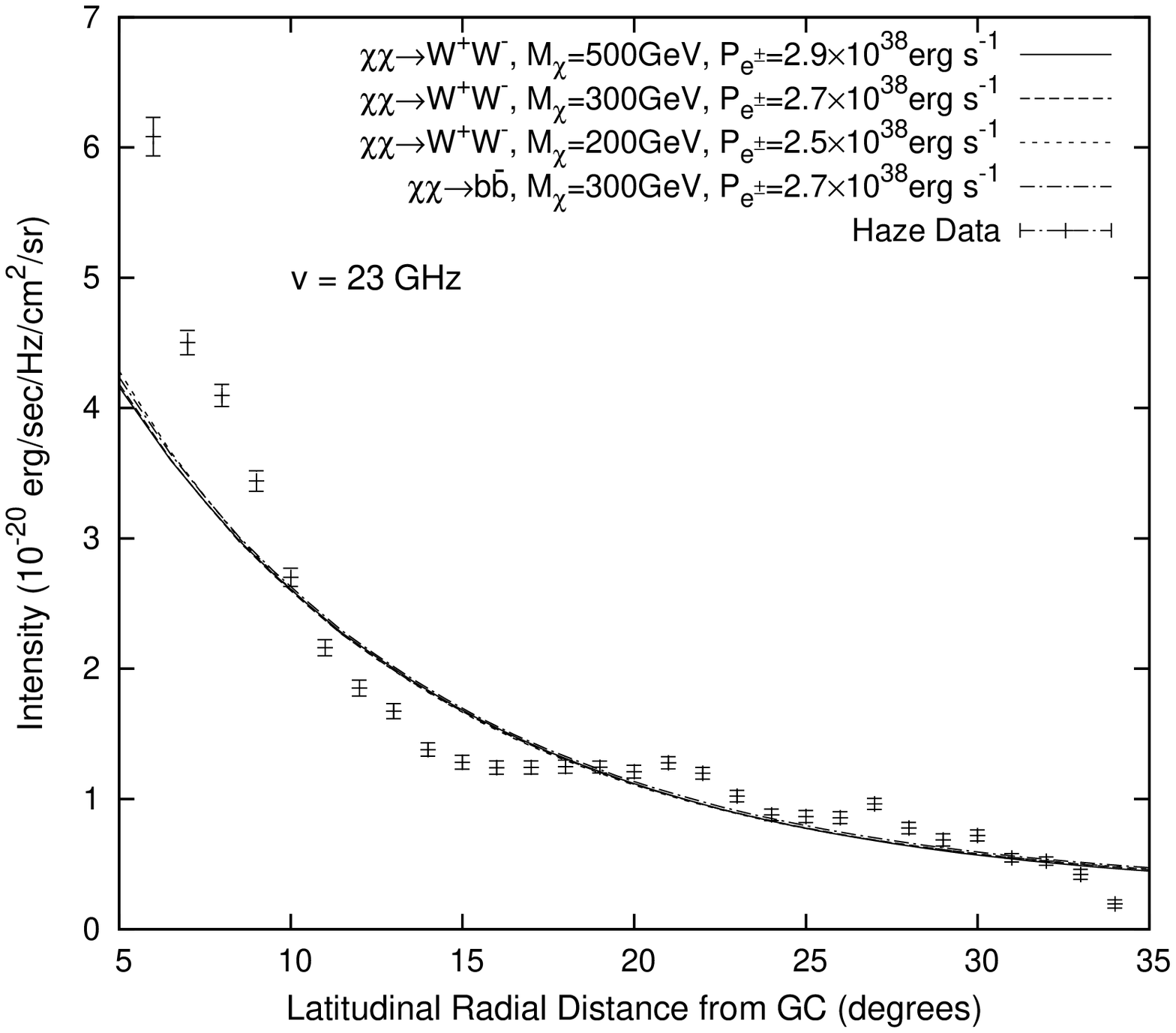, scale=0.45}
\epsfig{figure = 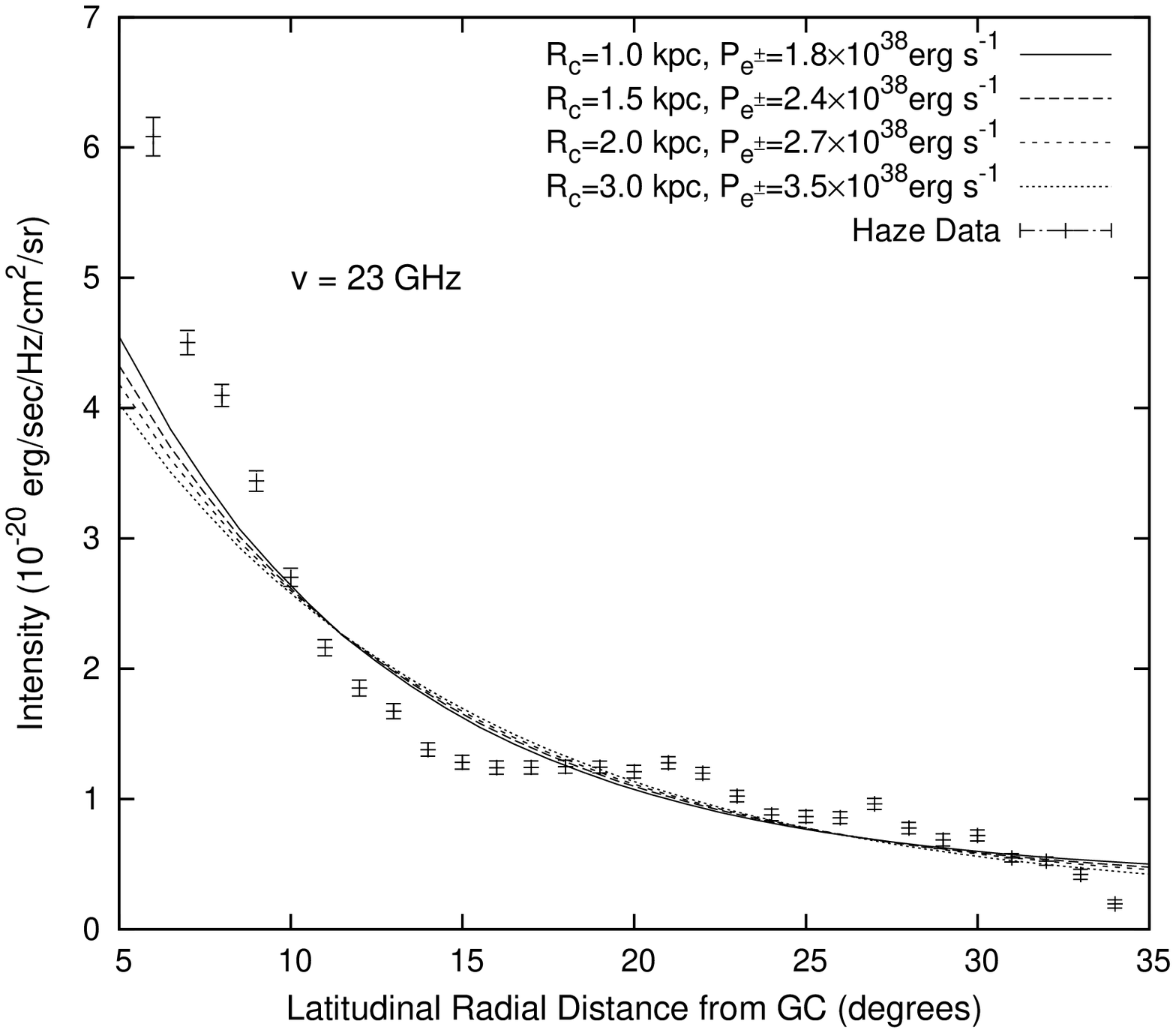, scale=0.45}\\
\epsfig{figure = 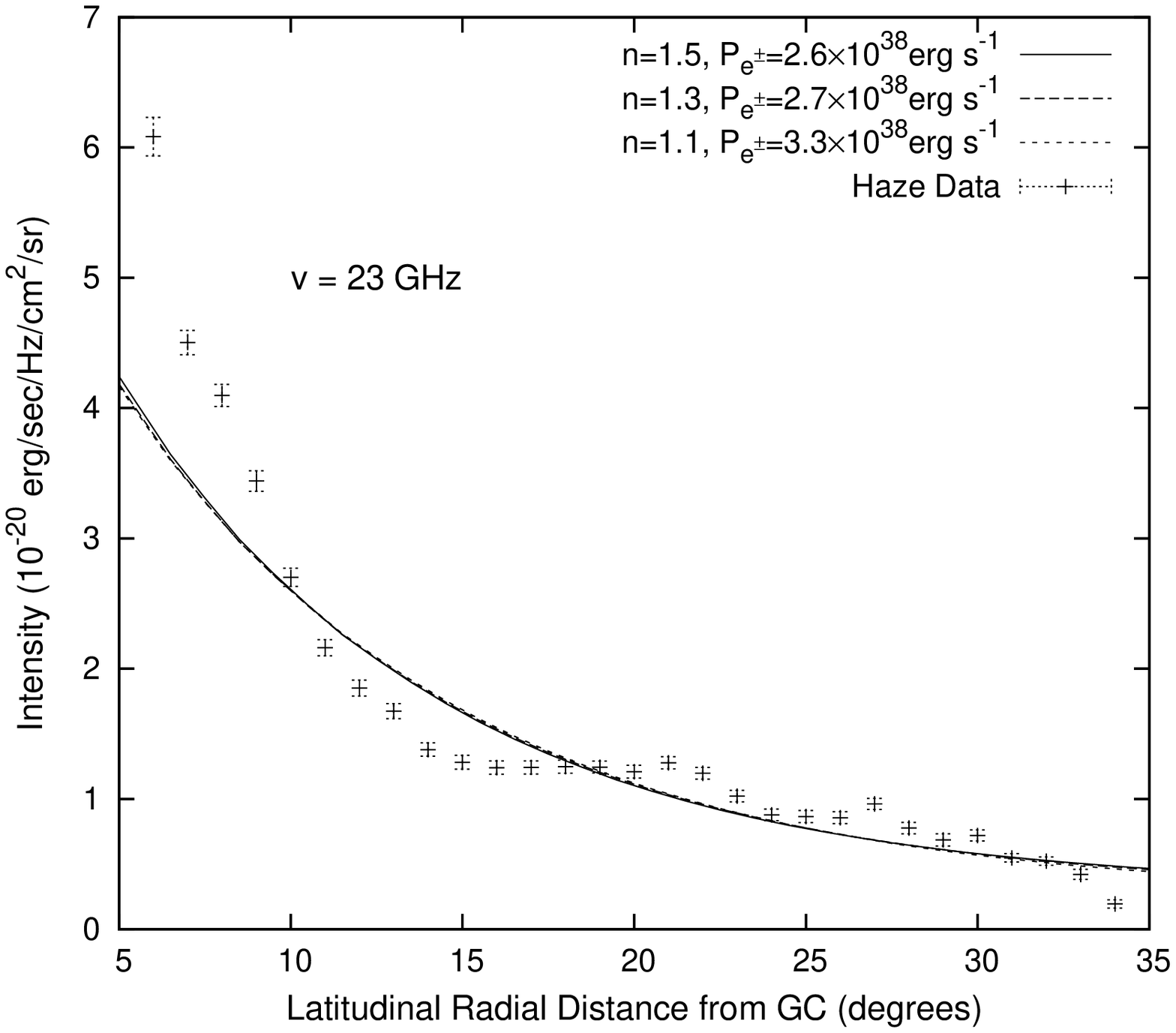, scale=0.45}
\epsfig{figure = 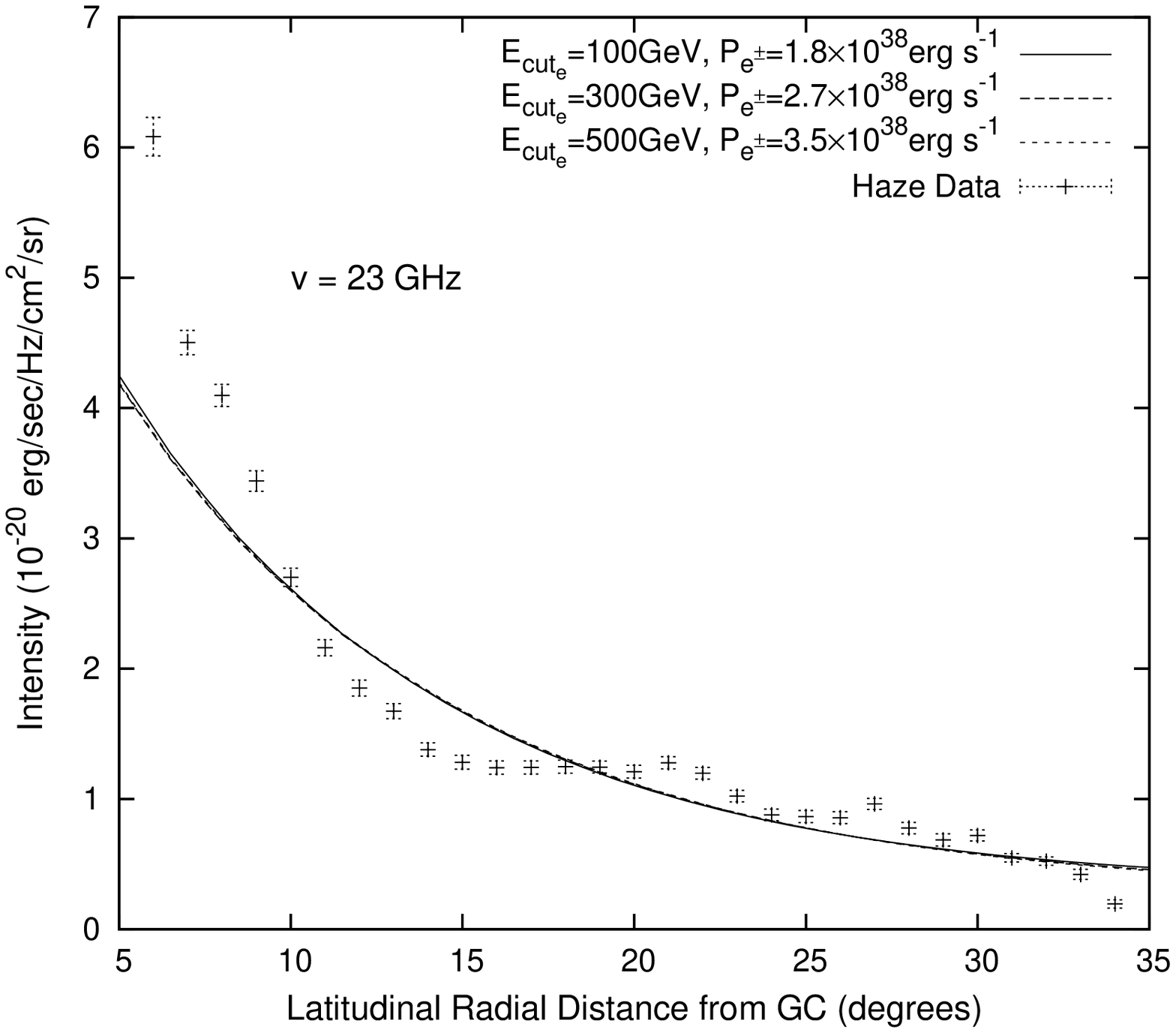, scale=0.45}
\end{center}
\vspace{-8mm}
\noindent
\caption{\small 
Fits to {\it WMAP} microwave haze data from Milky Way halo MSPs and DM 
with the same parameters as in Figure~\ref{VaryParsFermi}.
}
\label{VaryParsWMAP}
\vspace{1mm}
\end{figure}

In Figures \ref{VaryParsFermi} and \ref{VaryParsWMAP},
we address the question whether the
observationally unknown parameters used in Figure \ref{GammaHaze} were fine tuned.
In particular, we study the dependence 
on the index and cutoff of $e^+e^-$ injection spectrum from MSPs,
on the scaling radius of the MSP distribution $R_{\rm c}$,
on the DM mass, and on the DM annihilation 
channel%
\footnote{
We show two cases of annihilation channels
$\chi\chi \rightarrow W^{+}W^{-}$ and
 $\chi\chi \rightarrow b\bar{b}$.
The annihilation channel 
$\chi\chi \rightarrow \tau^{+}\tau^{-}$ has a similar $\gamma$-ray spectrum
to $\chi\chi \rightarrow W^{+}W^{-}$ 
but the number of high energy photons is lower by about a factor of 4.
The channel $\chi\chi \rightarrow ZZ$ 
produces a softer $\g$-ray spectrum than $\chi\chi \rightarrow W^{+}W^{-}$;
as a result, a higher $M_{\rm DM}$ will be needed to account 
for the high energy part of the $\gamma$-ray haze.
}.
We show that rather large changes of these parameters do not change our conclusions.

In this paper we have studied possible contributions to the $\g$-ray {\it Fermi} haze.
We have shown that two major contributions may come from the millisecond pulsars and
annihilating dark matter.
The electron and positron emission from MSPs is uncertain and we considered two cases.
In the first case, we assume that a large part of MSP spin-down luminosity is
transformed in the $e^+e^-$ flux.
In this case, a population of $3\times 10^4$ MSPs in the Milky Way halo
can explain the $\g$-ray haze 
if about $50\%$ of the MSPs spin-down is transformed into $e^+e^-$ 
and about $10\%$ is transformed in pulsed $\g$-rays.
The synchrotron emission from MSP electrons is consistent with the 
{\it WMAP} microwave haze data as well.
Prompt $\g$-ray emission from a DM particle annihilating into $W^+W^-$ 
contributes significantly around 100 GeV.
In the second case, we assumed that practically no MSP spin-down energy goes into $e^+e^-$.
The $\g$-ray haze can be explained by the pulsed $\g$-rays from MSPs and 
a DM particle annihilating into $W^+W^-$ with a boost factor 3.
However, the synchrotron emission (of $e^+e^-$ from our assumed DM model) 
is insufficient to explain the microwave haze data.
Another alternative to explain the $\g$-haze in the case of small $e^+e^-$ emission from MSPs
is DM particles with leptonic annihilation modes and
boost factors of order 100 together with pulsed $\g$-rays from MSPs.
The last scenario is also consistent with the {\it WMAP} microwave haze data.

In general, the $\g$-ray haze, if confirmed, provides a strong evidence for new sources
of $\g$-rays additional to the standard ones,
such as the $\pi^0$ production by cosmic rays and the ICS photons from
the electrons accelerated by SNe in the Galactic plane.
Possible new sources of $\g$-rays include DM annihilation
and a large population of MSPs in the stellar halo of the Milky Way. 
The $\g$-ray haze may also suggest a recent AGN activity 
or an existence of a bipolar Galactic wind \citep{2010ApJ...724.1044S}.

\bigskip
\bigskip

\newpage
\noindent
{\large \bf Acknowledgments.}

\noindent
The authors are thankful to Greg Dobler,
Claude-Andr\'e Faucher-Gigu\`ere, Douglas Finkbeiner,
Andrei Gruzinov, Mario Juric,
Daniel Kasen, Andrew MacFadyen, Neal Weiner, and Adi Zolotov
for valuable discussions and comments.
This work is supported in part 
by the Russian Foundation of Basic Research under grant no. RFBR 10-02-01315 (D.M.), 
by the NSF grant no. PHY-0758032 (D.M.),
by DOE OJI grant no. DE-FG02-06E R41417 (I.C.), 
by the Mark Leslie Graduate Assistantship (I.C.),
and by the NSF Astronomy and Astrophysics Postdoctoral Fellowship under grant no. AST-0702957 (J.D.G.).

% end body
%:

\bibliography{HazePapers}         %or whatever your .bib file is

\end{document}